\documentclass[useAMS,usenatbib]{mn2e}
\usepackage{times}
\usepackage{graphicx}

\newif\ifAMStwofonts
\AMStwofontstrue

\newcommand{\kms}{\,km\,s$^{-1}$}
\newcommand{\simgt}{\lower.5ex\hbox{$\; \buildrel > \over \sim \;$}}
\newcommand{\simlt}{\lower.5ex\hbox{$\; \buildrel < \over \sim \;$}}
\newcommand{\Msun}{M$_\odot$}

\title[Variation of IMF slope with velocity dispersion]{Systematic
  variation of the stellar Initial Mass Function with velocity
  dispersion in early-type galaxies}

\author[Ferreras et al.]{Ignacio Ferreras$^1$, Francesco La Barbera$^2$, 
Ignacio G. de la Rosa$^{3,4}$, Alexandre Vazdekis$^{3,4}$,
\newauthor
Reinaldo R. de Carvalho$^5$, Jes\'us Falc\'on-Barroso$^{3,4}$, 
Elena Ricciardelli$^6$\\
$^1$ Mullard Space Science Laboratory, University College London, 
Holmbury St Mary, Dorking, Surrey RH5 6NT \\
$^2$ INAF - Osservatorio Astronomico di Capodimonte, Napoli, Italy \\
$^3$ Instituto de Astrof\'\i sica de Canarias, C/ V\'\i a L\'actea
s/n, La Laguna, E-38200 La Laguna, Tenerife, Spain\\ 
$^4$ Departamento de Astrof\'\i sica,
Universidad de La Laguna, E-38205 La Laguna, Tenerife, Spain \\
$^5$ Instituto Nacional de Pesquisas Espaciais/MCT, S.J. dos Campos, Brazil\\
$^6$ Departament d'Astronomia i Astrofisica, Universitat de
  Valencia, C/Dr Moliner 50, E-46100, Burjassot, Valencia, Spain}

\begin{document}
\date{\sl MNRAS Letters: Accepted 2012 October 11. Received 2012 October 11; in original form 2012 August 15}
\pagerange{\pageref{firstpage}--\pageref{lastpage}} \pubyear{2012}
\maketitle
\label{firstpage}

\begin{abstract}
An essential component of galaxy formation theory is the stellar
initial mass function (IMF), that describes the parent distribution of
stellar mass in star forming regions. We present observational
evidence in a sample of early-type galaxies (ETGs) of a tight
correlation between central velocity dispersion and the strength of
several absorption features sensitive to the presence of low-mass
stars. Our sample comprises $\sim 40,000$ ETGs from the SPIDER survey
(z$\simlt$0.1).  The data -- extracted from the Sloan Digital Sky
Survey -- are combined, rejecting both noisy data, and spectra with
contamination from telluric lines, resulting in a set of 18 stacked
spectra at high signal-to-noise ratio ($S/N\simgt 400$\AA$^{-1}$). A
combined analysis of IMF-sensitive line strengths and spectral fitting
is performed with the latest state-of the art population synthesis
models (an extended version of the MILES models).  A significant trend
is found between IMF slope and velocity dispersion, towards an excess
of low-mass stars in the most massive galaxies.  Although we emphasize
that accurate values of the IMF slope will require a detailed analysis
of chemical composition (such as [$\alpha$/Fe] or even individual
element abundance ratios), the observed trends suggest that low-mass
ETGs are better fit by a Kroupa-like IMF, whereas massive galaxies
require bottom-heavy IMFs, exceeding the Salpeter slope at
$\sigma\simgt 200$\kms\,.
\end{abstract}


\section{Introduction}
\label{sec:intro}

The Initial Mass Function (hereafter IMF) is defined as the
distribution of stellar masses in a single population at the time of
birth. It has been usually considered a universal function, partly
because of the complexities in obtaining proper observational
constraints. The traditional approximation by a single power law
\citep{Salpeter:55} has undergone numerous updates, with more complex
functions that include a significant flattening of the slope towards
low-mass stars \citep{MillerScalo:79,Scalo:86,Kroupa:01,Chabrier:03}.
For a recent review on the IMF and its possible variations, see
\citet{Bastian:10}. 

\begin{figure}
\begin{center}
\includegraphics[width=75mm]{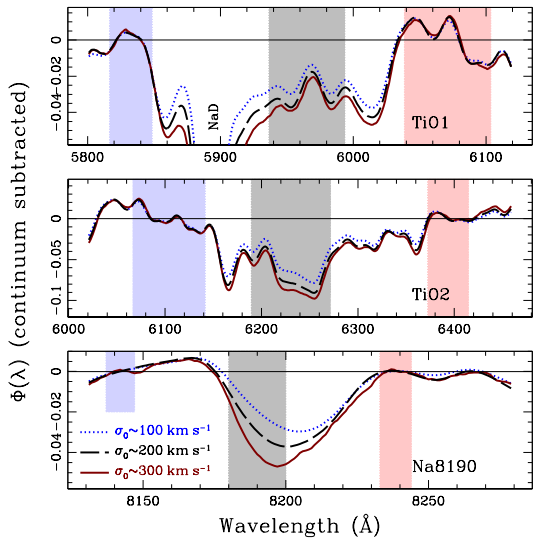}
\caption{Trend of the three absorption features targeted in this paper
  (TiO1, TiO2 and Na8190) with respect to central velocity dispersion,
  for stacked SDSS spectra of ETGs from the SPIDER sample. The SEDs
  have been smoothed to a common velocity dispersion of 300\kms\, with the
  spectral resolution of SDSS, and
  continuum subtracted with a second order polynomial. 
  We only show, for clarity, three out of the 18 stacks, spanning the full
  range of velocity dispersion. The shaded regions mark the positions
  of the red and blue sidebands, and the central bandpass.
\label{fig:stacks}
}
\end{center}
\end{figure} 

Past analyses based on dynamical modelling favoured Kroupa-like IMFs
in ETGs \citep[][]{Cappellari:06}.  This result was also found with
strong gravitational lensing on the bulge of a late-type galaxy
\citep{Ferreras:10}.  However, the IMF was suspected to depend on
galaxy mass, when extending the analysis to more massive systems, as in
the dynamical modelling of \citet{Thomas:11}, with galaxy morphology
also playing a role \citep{Brewer:12}. Constraining the low-mass
portion of the IMF is a challenging task, as faint, low-mass stars do
not contribute significantly to the integrated spectrum. The first
observational attempts to constrain the low-mass end of the IMF were
made by \citet{Cohen:78} and \citet{FaberFrench:80}, towards the
centers of M31 and M32, measuring several spectral features sensitive
to the relative contribution of M dwarves and M giants. Later,
\cite{Carter:86} extended the study to a sample of massive early-type
galaxies, and found that NaI was enhanced, especially in massive
galaxies, with strong radial gradients (being most enhanced in the
central region). More recently, \citet{Cenarro:03} proposed a relation
towards an excess of low-mass stars in massive galaxies, from a study
of the Ca{\sc II} triplet region at
$\sim$8500\,\AA\,. \citet{VdKConroy:10} probed two features in the NIR
that are strongly dependent on the fraction of low-mass stars, to find
a bottom-heavy IMF in a sample of four Virgo ETGs.  This result was
recently extended to an additional set of 34 ETGs from the SAURON
survey, with similar conclusions \citep{VdKConroy:12}, favouring the
interpretation of a non-universal IMF. An alternative scenario to
explain these data would invoke over/under abundances of individual
elements \cite[see e.g.][]{Worthey:98}.  Bottom-heavy IMFs would imply
that massive galaxies observed at high redshift are even more massive
than the standard assumption of a Kroupa-like IMF, posing challenging
constraints on the physics underlying the conversion of gas into stars
during the first stages of galaxy formation.  This issue is of special
relevance to the standard paradigm of hierarchical galaxy formation,
whereby massive galaxies are supposed to grow mainly through the
merging of smaller systems, with the growth rate limited by cosmology
\citep[see, e.g.,][]{DeLucia:06}, requiring alternative channels of
massive galaxy formation, such as cold accretion \citep[see,
  e.g.][]{BirnDek:03}.

Simple theories of star formation suggest a power law behaviour, with
a truncation at the low-mass end, around 0.3\,\Msun
\citep{Larson:05,McKee:07}, consistent with some of the most popular
IMFs used \citep{Kroupa:01,Chabrier:03}. However, due to the lack of
strong observable tracers in the observed spectra arising from low
mass stars, a robust constraint at the low-mass end of the IMF has
remained elusive for years. In addition to the work cited above, more
recent evidence has been accumulating
\citep{VdKConroy:10,Spiniello:12,Smith:12}, towards the possibility of
a non-universal IMF. This hypothesis is further supported by an
independent study that made use of detailed dynamical modelling of
nearby spheroidal galaxies \citep{Cappellari:12}. Strong gravitational
lensing over galaxy scales also hint at bottom heavy IMFs in massive
galaxies \citep{Auger:10}. The implications of a change in the IMF
warrant further analysis of this problem.

\begin{figure}
\begin{center}
\includegraphics[width=75mm]{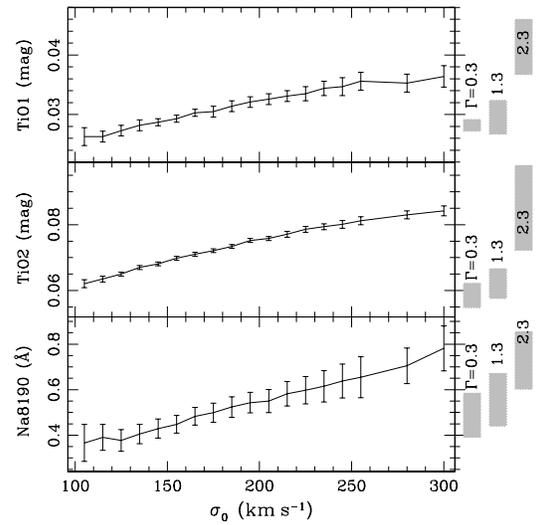}
\caption{Trend of the equivalent widths of TiO1, TiO2 and Na8190, with
  respect to velocity dispersion. All measurements are performed on
  data convolved to a common velocity dispersion of
  300\,\kms\ with the spectral response of SDSS. The error bars give
  the statistical error from the stacks, at the $3\,\sigma$ level. The
  shaded regions on the right correspond to model predictions for
  three choices of IMF unimodal slope, as labelled, spanning a range
  of ages from $5$ to $10$\,Gyr and metallicity from $\log
  Z/Z_\odot=-0.4$ to $+0.2$
  \label{fig:EWs}}
\end{center}
\end{figure} 

\section{Sample}
\label{sec:sample}

The hypothesis of a systematic change in the Initial Stellar Mass
Function with respect to global properties of galaxies, such as mass
or luminosity, are optimally tested on samples of galaxies that
harbour stellar populations as homogenous as possible, for a clean
analysis of the spectra. The SPIDER sample \citep[Spheroids
  Panchromatic Investigation in Different Environmental
  Regions,][]{SpiderI} is optimal for this purpose, as it comprises
39,993 nearby (0.05$<$z$<$0.095) ETGs, in a wide range of masses, from
$2\times 10^{10}$ to $10^{12}$\Msun \citep[dynamical
  mass,][]{delaRosa:12}. Spectra of these galaxies were retrieved from
the sixth data release of the Sloan Digital Sky Survey
\citep{SDSS:DR7} (SDSS), de-redshifted to a common rest-frame (with
1\AA\ binning) and corrected for foreground Galactic extinction
\citep[see ][for details]{SpiderI}.  In order to test variations of
the IMF, we assemble 18 stacked spectra spanning over a wide range of
central velocity dispersions ($\sigma$) between 100 and 320\kms\ in
steps of 10\kms, except for the two bins for the highest values of
$\sigma$, where -- because of the smaller number of galaxies in the
sample -- we adopted the range [260,280], and [280,320]\kms\,,
respectively.  We select the sample according to $\sigma$, because (i)
the underlying stellar populations of ETGs are known to correlate
strongly with this observable \citep{Bernardi:05}, and (ii) the
spectral analysis depends sensitively on velocity dispersion, as a
higher $\sigma$ introduces an effectively lower spectral resolution,
an issue that we take into account for a consistent comparison (see
below).

Given the redshift range of the SPIDER sample, the 3~arcsec diameter
of the SDSS fibers maps the central $\sim 2-3.5\,h^{-1}$\,kpc.  For
each bin, we perform median stacking of the available spectra,
considering only pixels with no SDSS flag raised.
In order to avoid possible biases related to intrinsic dust and
differences in signal to noise ratio ($S/N$) within the bin, we
excluded spectra in the lowest quartile of the $S/N$ distribution, as
well as those galaxies ($14 \%$ of the entire sample) with $E(B-V) >
0.1$~mag. The colour excess is measured by fitting each SDSS spectrum
with a variety of stellar population models, assuming
a~\citet{Cardelli} extinction law, as detailed in~\citet{Swindle}.

\section{IMF-sensitive indicators}
\label{sec:IMF}

The SDSS spectral range allows us to study several IMF-sensitive
spectral features, such as the Na\,I doublet at $\lambda\lambda
8183,8195$\,\AA\ \citep[][]{SchiavonNaD:97}.  This feature is
prominent in the atmospheres of low-mass dwarf stars.  Therefore, its
strength in a stellar population depends quite sensitively on the
shape of the IMF at the low-mass end. Over the IMF-sensitive
wavelength range (8183--8195\AA), our stacked spectra feature a
remarkably high $S/N$, with an average of $\sim 400$ per \AA\, for the
lowest and highest $\sigma$ bins, reaching values as high as $S/N\sim
850$ per \AA\ in the stack with velocity dispersion $\sigma \sim
160$\,\kms.  We use here a slightly modified version of the NaI8200A
index proposed by \citet{MIUSCAT:12}, defined as Na8190, where the
blue passband is measured in the [8137,8147]\AA\ interval (instead of
[8164,8173]\AA), where the MIUSCAT models are found to give a better
match to the data, within $\sim 0.3 \%$ (compared to $1 \%$ at
$\lambda \sim 8170$\AA\,). The red and central bandpasses are the same
as for the NaI8200A index.

We performed extensive tests to check for possible contamination of
the Na8190 feature from sky lines, by combining (1) all galaxies at
each $\lambda$ regardless of the SDSS flag value; (2) only flux values
not affected by sky absorption (i.e.  1\AA\ apart from any telluric
line) and sky emission; and (3) only ETGs at the lowest redshifts in
the SPIDER sample ($0.05\! \le\!  z\!  \le\! 0.07$), where the Na\,I
doublet is observed in a region free from any significant sky
contamination.  For all $\sigma$ bins, the variation of the average
Na\,I flux, using the above three stacking procedures, was found to be
smaller than $0.06 \%$, i.e.  negligible for our purposes.

In order to increase the robustness of the analysis against possible
biases regarding an overabundance of [Na/Fe] in massive galaxies, we
also consider two line strengths targeting TiO, a further discriminant
with respect to the presence of low-mass stars.  We use the TiO1 and
TiO2 definitions from the standard Lick system \citep[see
  e.g.][]{Trager:98}. The TiO2 index has been recently used by
\citet{Spiniello:12} to constrain the IMF of a number of early-type
gravitational lenses. Fig.~\ref{fig:stacks} shows three of our 18
stacks, in the regions of interest for this work.  All SEDs have been
smoothed to a common velocity dispersion of 300\kms\ for a consistent
comparison, and were continuum subtracted, using a second order
polynomial. A significant trend is apparent from low- to high-velocity
dispersion of all indices. This trend encodes a complex range of
variations in the underlying stellar populations, most notably age,
metallicity, abundance ratios, and IMF.

Note in the Na8190 region ({\sl bottom}), the centroid of the line
shifts bluewards as $\sigma$ increases. However, because of the
asymmetry of the spectral region around the line, more flux would be
expected to contribute from the red side of the line, therefore
shifting the position of the centroid redwards. This shifting reveals
that it is the absorption of Na at 8190\AA\ -- and not other
contaminating lines in the vicinity -- that changes with respect to
the velocity dispersion of the ETG. The two most obvious
interpretations of this effect involve either an overabundance of [Na/Fe]
\citep{Worthey:98}, or a change in the IMF \citep{VdKConroy:10}. In a
forthcoming paper (La Barbera et al., in preparation) we explore in detail 
the effect of individual overabundances. However, in this letter, we
combine the three line strengths -- which rely on different
species -- to confirm the trend towards a bottom-heavy IMF in
massive galaxies. 

\begin{figure}
\begin{center}
\includegraphics[width=75mm]{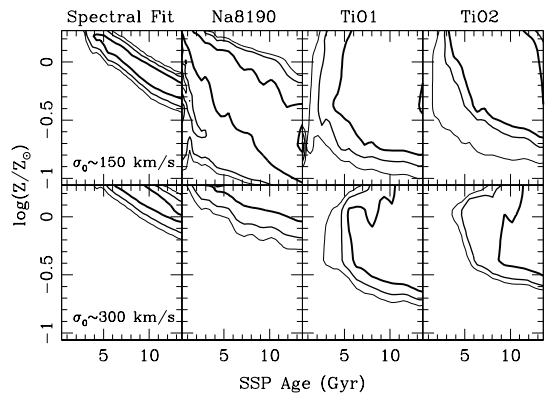}
\caption{The 68, 95 and 99\% confidence levels of the likelihood
  from {\sl individual} constraints (from left to right) of spectral
  fitting, Na8190, TiO1 and TiO2 are shown on the age-metallicity
  plane for two of the stacks, as labelled by their central
  velocity dispersion. Bimodal IMFs are used here, although the
results for the unimodal case are similar.
\label{fig:SpecFit}
}
\end{center}
\end{figure} 

Fig.~\ref{fig:EWs} shows the line strengths of the stacked SEDs, for
the three IMF-sensitive indices. The strengths are corrected to a
common broadening of $\sigma_{\rm ref}=$300\kms (in addition to the
wavelength-dependent SDSS resolution). A strong correlation of these
line strengths with velocity dispersion is evident.  We checked that
the fixed aperture of the fibres used by SDSS to retrieve spectra did
not introduce a bias with respect to size, by comparing the change in
line strengths with respect to the angular extent of the effective
radius. The variation of the indices, at fixed $\sigma$, is around an
order of magnitude smaller than the trend shown in
Fig.~\ref{fig:EWs}. The shaded regions on the right of the figure
motivate the methodology followed in this paper. Each one gives the
model predictions for a choice of unimodal IMF slope (labelled), over
a range of SSP ages and metallicities.  The values of all three
indices at high velocity dispersion can only be reconciled with a
bottom-heavy IMF. Our methodology -- explained below -- consists of
removing the degeneracies from age and metallicity by combining line
strength information with spectral fitting over the optical range.

\section{Constraining the IMF}
\label{sec:analysis}

We make use of MIUSCAT \citep{MIUSCAT:12,MIUSCAT2:12}, the spectrally
extended version of the stellar population synthesis models MILES
\citep{MILES:10}, in order to map the systematic trend of the TiO1,
TiO2 and Na8190 line strengths with respect to IMF slope. These models
combine state-of-the art stellar libraries in the optical and NIR
windows, creating a set of spectra with a uniform resolution of
2.51\AA\ throughout the wavelength range 3465--9469\AA\ . Our data are
compared with grids of simple stellar populations (SSPs), assuming
either a unimodal or a bimodal IMF \citep[as defined in][]{vaz:96},
over a stellar mass range $0.1-100$\,\Msun.  For the unimodal case, we
adopt the logarithmic slope $\Gamma = x - 1$, where $dN/dm\propto
m^{-x}$ is the Initial Mass Function, such that the
\citet{Salpeter:55} IMF corresponds to $\Gamma=1.35$. A bimodal IMF
replaces the $M<0.6M_\odot$ interval by a flat portion and a spline to
match the power law at the high mass end. It gives a closer
representation of Kroupa-like IMFs for $\Gamma=1.3$ \citep[see,
  e.g. Figure~1 of ][]{vaz:03}. We use MIUSCAT SSPs spanning a wide
range of ages (1--13~Gyr) and metallicities ($-1.0 \! \le \! \log
Z/Z_\odot \!  \le \!  +0.22$), for different values of $\Gamma=\{0.3,
0.8, 1.0, 1.3, 1.5, 1.8, 2.0, 2.3, 2.8, 3.3\}$.  In order to obtain an
accurate estimate of $\sigma$ for each stack, we perform spectral
fitting with the software STARLIGHT \citep{Cid:05}, in the range
3800--8400\,\AA.  STARLIGHT can be used to extract full star formation
histories \citep[see e.g.][]{delaRosa:12}, but we are only interested
here in assessing the robustness of the measured $\sigma$ with respect
to the basis SSPs.  We find no significant trend when choosing
template SSPs with different IMFs, with a variation
$\Delta\sigma\simlt $1\,\kms\,. Furthermore, for each stack, the
$\sigma$ determined by STARLIGHT is consistent, within $\sim 10 \%$,
with the median $\sigma$ of the stacked spectra from SDSS.

We parameterise a stellar population by an age, metallicity, and IMF
slope. More detailed models will be presented elsewhere (La Barbera et
al., in preparation). We note that, according to \citet{TMJ:11}, both
TiO1 and TiO2 are rather insensitive to non-solar abundance
ratios. Furthermore, \citet{CvdK:12} find an anticorrelation between
[$\alpha$/Fe] and NaI0.82 strength (similar to our Na8190
index). Hence, we do not expect our results, regarding the
non-universality of the IMF, to be affected by the well-known
overabundance of [$\alpha$/Fe] in massive galaxies \citep[see
  e.g.][]{Trager:00}. To confirm this point, we studied two subsamples
within the highest-$\sigma$ bin, split with respect to the
distribution of [$\alpha$/Fe] values, constructing low-($\sim +0.1$)
and high-($\sim +0.3$) [$\alpha$/Fe] stacks. The measured
TiO1(mag)/TiO2(mag)/Na8190(\AA) strengths are $0.0351\pm 0.0014$ /
$0.0860\pm 0.0010$ / $0.74\pm 0.07$, for the high [$\alpha$/Fe]
subsample, $0.0352\pm 0.0015$ / $0.0833\pm 0.0010$ / $0.77\pm 0.08$,
for the low [$\alpha$/Fe] subsample, which is fully consistent with
the original stack, including all values of [$\alpha$/Fe]: $0.0364\pm
0.0006$ / $0.0842\pm 0.0005$ / $0.78\pm 0.03$.

Our analysis follows a hybrid method, whereby two independent
probability distribution functions for age, metallicity and IMF slope
are obtained using spectral fitting and line strengths,
respectively. For the former, we fit the region 3900--5400\AA\ , using
SSPs from MIUSCAT, after convolving them from the original
2.51\AA\ resolution, to the fiducial velocity dispersion of 300\,\kms
, plus SDSS spectral resolution.  A grid of $32\times32\times 10$
models in age, log Z and $\Gamma$ is used to obtain a probability
distribution function (PDF), via the likelihood ${\cal
  L}\propto\exp(-\Delta\chi^2/2)$. For each line strength, we also
define a second PDF in the same grid. The joint PDF is defined as the
product of the corresponding (independent) PDFs.  Finally, we
marginalize over all parameters but $\Gamma$, to obtain the PDF of the
IMF slope. Fig.~\ref{fig:SpecFit} illustrates the constraining power
of the different observables, individually, on the age and metallicity
of the model SSP. Notice that the combined analysis of the hybrid
method is acceptable for all three cases (i.e.  the joint likelihood
does not reduce to the product of mutually unlikely regions of
parameter space).

\begin{figure}
\begin{center}
\includegraphics[width=75mm]{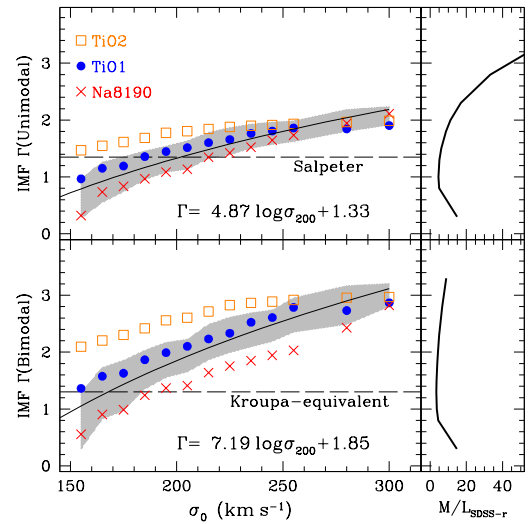}
\caption{Variation of the IMF slope -- unimodal ({\sl top}) and
  bimodal ({\sl bottom}) distributions -- against central velocity
  dispersion. The shaded region corresponds to the 68\% confidence
  level of the joint PDF including spectral fitting and all three line
  strengths (TiO1, TiO2 and Na8190). The Salpeter (unimodal) and
  Kroupa-equivalent (bimodal) cases are shown as horizontal dashed
  lines. A least squares fit to the data is shown for reference
  (solid line). The rightmost panels give the stellar mass-to-light
  ratios in the SDSS-$r$ band for a 10\,Gyr old population at solar
  metalliticy, illustrating the large variations one could expect
  depending on the choice of IMF.
\label{fig:IMF}
}
\end{center}
\end{figure} 

\section{Conclusions}
\label{sec:conclusions}

Fig.~\ref{fig:EWs} motivates the need to invoke a non-universal IMF
when comparing galaxies over a range of velocity dispersions, with an
increased contribution from low-mass stars in the most massive
galaxies.  Fig.~\ref{fig:IMF} presents the best-fit slope of the IMF,
when marginalizing over age and metallicity, as a function of velocity
dispersion, given by the probability weighted estimates using the
joint PDF, for three cases, depending on whether TiO1 (filled
circles), TiO2 (open squares), or Na8190 (crosses) are used in the
analysis.  Both unimodal ({\sl top}) and bimodal ({\sl bottom}) IMFs
are considered. We note that at low velocity dispersion, the
constraint on IMF slope becomes more challenging, as low-mass ETGs
feature complex star formation histories \citep[see
  e.g.][]{delaRosa:12}, making the SSP approach used here not fully
applicable. Therefore, Fig.~\ref{fig:IMF} only shows the range
150--300\,\kms, where the approximation of a SSP is justified. The
general trend towards a bottom-heavy IMF is evident in all cases.  The
result for the joint PDF corresponding to all three indices plus
spectral fitting is shown as a grey shaded region -- extending over
the 68\% confidence level -- with a black line giving a simple least
squares fit to the data.  We emphasize here that accurate values of
the IMF slope will require a detailed analysis of abundance ratios
(such as, e.g., [$\alpha$/Fe] or [Na/Fe]).  The predicted IMF slopes
for massive ETGs do not pose a problem to optical-NIR photometry,
where the contribution from low-mass stars would be most important. As
an example, the MIUSCAT models for a 10\,Gyr population at solar
metallicity give $V-K=3.04$ for a (bimodal) IMF slope $\Gamma=0.30$,
versus $V-K=3.03$ for $\Gamma=2.80$ (A unimodal distribution will give
an excess in $V-K$ of $\sim 0.25$\,mag over a similar range of
$\Gamma$). Hence, a variation in the IMF is compatible with the
observed $V-K$ colours. Furthermore, $V-K$ colours alone cannot be
used to constrain the IMF, a well-known result, that explains why
constraints on the IMF have remained elusive.  The $V-K$ colour of our
sample ranges from $3.0$ at the lowest velocity dispersion bin, to
$3.3$ at a velocity dispersion of 300\,\kms \citep{SpiderI}.

The rightmost panels of Fig.~\ref{fig:IMF} should serve as a warning
to applications of these trends to infer stellar masses. The M/L in
the SDSS-r band is shown for a typical old, metal-rich population,
such as those found in these galaxies. The difference between a
unimodal ({\sl top}) and a bimodal ({\sl bottom}) IMF is very
significant for massive galaxies, whereas the fits for either choice
of IMF are equally acceptable.  The correlation presented in this
letter is consistent with studies based on different methods,
involving strong gravitational lensing, or dynamical modelling
\citep[see, e.g.][]{Treu:10,Thomas:11,Cappellari:12}.  Regarding the
study of stellar populations, our work is consistent with the trend in
\citet{Cenarro:03}, based on a spectroscopic analysis of the Ca{\sc
  II} triplet region. In addition, the recent {\sl photometric}
analysis of \citet[][see their Figure~16]{MIUSCAT2:12}, is compatible
with our analysis.  Our results are also in agreement with the
findings of \citet{VdKConroy:10,VdKConroy:12}. We emphasize that our
analysis is based on a completely independent stellar library to their
work. For instance, in the region around the Na8190 feature, the
MIUSCAT models use stellar spectra from the Indo-US library
\citep{IndoUs}, whereas the models of \citet{CvdK:12} use the IRTF
library \citep{IRTF1,IRTF2}. In a forthcoming paper, we will explore
in detail different aspects related to the methodology, including the
contribution from enhacements of individual elements, non-solar
[$\alpha$/Fe], or composite stellar populations. The large size of our
sample -- comprising nearly 40,000 ETGs -- allowed us to go beyond a
simple test of non-universality of the IMF, enabling us to obtain a
trend between the IMF slope and velocity dispersion in more detail
than previously found. The existing correlation between $\Gamma$ and
velocity dispersion in ETGs suggests a significant transition in the
properties of star forming regions as a function of galaxy mass, an
issue with important implications on galaxy formation theories.

\bigskip

IF would like to acknowledge the hospitality of INAF/OAC and the
IAC. JFB acknowledges support from the Ram\'on y Cajal programme by the
Spanish Ministry of Economy and Competitiveness (MINECO). This work
has been supported by the Programa Nacional de Astronom\'\i a y
Astrof\'\i sica of MINECO,
under grants AYA2010- 21322-C03-01 and AYA2010-21322-C03-02
and by the Generalitat Valenciana (grant PROMETEO-2009-103).  This
letter made use of data from the SDSS
(http://www.sdss.org/collaboration/credits.html), and also from the
UKIRT Infrared Deep Sky Survey~\citep{Law:07}.


\label{lastpage}

\begin{thebibliography}{}
\bibitem[Abazajian et al. (2009)]{SDSS:DR7} 
Abazajian, K.~N., et al., 2009, ApJS, 182, 543

\bibitem[Auger et al. (2010)]{Auger:10} 
Auger, M.~W., Treu, T., Gavazzi, R., Bolton, A.~S., 
Koopmans, L.~V.~E. \& Marshall, P.~J., 2010, ApJ, 721, L163

\bibitem[Bastian et al. (2010)]{Bastian:10}
Bastian, N., Covey, K.~R. \& Meyer, M.~R. 
2010, ARA\&A, 48, 339

\bibitem[Bernardi et al. (2005)]{Bernardi:05} 
Bernardi, M., Sheth, R.~K., Nichol, R.~C.,
Schneider, D.~P. \& Brinkmann, J.
2005, AJ, 129, 61

\bibitem[Birnboim \& Dekel (2003)]{BirnDek:03} 
Birnboim, Y. \& Dekel, A., 2003, MNRAS, 345, 349

\bibitem[Brewer et al. (2012)]{Brewer:12} 
Brewer, B.~J., et al. 2012, MNRAS, 422, 3574

\bibitem[Cappellari et al. (2006)]{Cappellari:06} 
Cappellari, M., et al. 2006 MNRAS, 366, 1126

\bibitem[Cappellari et al. (2012)]{Cappellari:12} 
Cappellari, M. et~al. 2012, Nature, 1202.3308

\bibitem[Cardelli, Clayton, Mathis (1989)]{Cardelli} 
Cardelli, J.~A., Clayton, G.~C., Mathis, J.~S., 1989, ApJ, 345, 245

\bibitem[Carter et al.(1986)]{Carter:86} 
Carter, D., Visvanathan, N. \& Pickles, A.~J., 1986, ApJ, 311, 637

\bibitem[Cenarro et al. (2003)]{Cenarro:03} 
Cenarro, A.~J., Gorgas, J., Vazdekis, A.,
Cardiel, N. \& Peletier, R.~F. 2003, MNRAS, 339, L12

\bibitem[Chabrier (2003)]{Chabrier:03}
Chabrier, G. 2003, PASP, 115, 763

\bibitem[Cid Fernandes et al. (2005)]{Cid:05}
Cid Fernandes, R., Mateus, A., Sodr\'e, L.,
Stasinska, G. \& Gomes, J.~M.,
2005, MNRAS, 358, 363

\bibitem[Cohen (1978)]{Cohen:78}
Cohen, J.~G., 1978, ApJ, 221, 788

\bibitem[Conroy \& van Dokkum (2012)]{CvdK:12}
Conroy, C. \& van~Dokkum, P.~G., 2012, ApJ, 747, 69

\bibitem[Cushing et al. (2005)]{IRTF1}
Cushing, M.~C., Rayner, J.~T., \& Vacca, W,~D., 2005, ApJ, 623, 1115

\bibitem[de la Rosa et al. (2012)]{delaRosa:12} 
de la Rosa, I.~G., la Barbera, F., Ferreras, I. \& 
de Carvalho, R.~R. 2012 MNRAS, 418, L74

\bibitem[De Lucia et al. (2006)]{DeLucia:06} 
De Lucia, G., Springel, V., White, S.~D.~M., 
Croton, D. \& Kauffmann, G.
2006, MNRAS, 366, 499

\bibitem[Faber \& French (1980)]{FaberFrench:80} 
Faber, S.~M. \& French, H.~B., 1980, ApJ, 235, 405

\bibitem[Ferreras et al. (2010)]{Ferreras:10} 
Ferreras, I., Saha, P., Leier, D.,
Courbin, F. \& Falco, E.~E. 2010, MNRAS, 409, L30

\bibitem[Kroupa (2001)]{Kroupa:01}
Kroupa, P., 2001, MNRAS, 322, 231

\bibitem[La Barbera et al. (2010)]{SpiderI}  
La Barbera, F., de Carvalho, R.R., de La Rosa, I.G., Lopes, P.A.A.,
Kohl-Moreira, J.L., Capelato, H.V., 2010, MNRAS, 408, 1313

\bibitem[Larson (2005)]{Larson:05} 
Larson, R.~B., 2005, MNRAS, 359, 211

\bibitem[Lawrence et al. (2007)]{Law:07} 
Lawrence, A., et al. 2007, MNRAS, 379, 1599

\bibitem[McKee \& Ostriker (2007)]{McKee:07}
McKee, C.~F. \& Ostriker, E.~C.
2007, ARA\&A, 45, 565

\bibitem[Miller \& Scalo (1979)]{MillerScalo:79} 
Miller, G.~E. \& Scalo, J.~M., 1979, ApJs, 41, 513

\bibitem[Rayner et al. (2009)]{IRTF2}
Rayner, J.~T., Cushing, M.~C. \& Vacca, W.~D., 2009, ApJS, 185, 289

\bibitem[Ricciardelli et al. (2012)]{MIUSCAT2:12}
Ricciardelli, E., Vazdekis, A., Cenarro, A.~J. \& Falc\'on-Barroso, J., 
2012, MNRAS, 424, 172

\bibitem[Salpeter (1955)]{Salpeter:55}
Salpeter, E.~E., 1955, ApJ, 121, 161

\bibitem[Scalo (1986)]{Scalo:86} 
Scalo, J.~M., 1986, Fund. Cosm. Phys., 11, 1

\bibitem[Schiavon et al. (1997)]{SchiavonNaD:97}
Schiavon, R., Barbuy, B., Rossi, S.~C.~F. \&
Milone, A. 1997, ApJ, 479, 902

\bibitem[Schiavon et al. (1997b)]{SchiavonFeH:97}
Schiavon, R., Barbuy, B. \& Singh, P.~D. 
1997b, ApJ, 484, 499

\bibitem[Smith et al. (2012)]{Smith:12}
Smith, R.~J, Lucey, J.~R. \& Carter, D., 2010, arXiv:1206.4311

\bibitem[Spiniello et al. (2012)]{Spiniello:12}
Spiniello, C., Trager, S.~C., Koopmans, L.~V.~E. \& 
Chen, Y., 2012, ApJ, 753, L32

\bibitem[Swindle et al.(2011)]{Swindle} Swindle, R., Gal, R.R., La Barbera, F., de Carvalho, R.R., 2011, AJ, 142, 118

\bibitem[Thomas et al. (2003)]{Thomas:03} 
Thomas, D., Maraston, C. \& Bender, R., 2003, MNRAS, 339, 897

\bibitem[Thomas et al. (2005)]{Thomas:05} 
Thomas, D., Maraston, C., Bender, R. \&
Mendes de Oliveira, C. 2005, ApJ, 621, 673

\bibitem[Thomas et al. (2011a)]{TMJ:11} 
Thomas, D., Maraston, C. \& Johansson, J.
2011a, MNRAS, 412, 2183

\bibitem[Thomas et al. (2011b)]{Thomas:11} 
Thomas, J., et al. 2011b, MNRAS, 415, 545

\bibitem[Trager et al. (1998)]{Trager:98} 
Trager, S.~C., Worthey, G., Faber, S.~M., Burstein, D. \& Gonz\'alez, J.~J.,
1998, ApJS, 116, 1

\bibitem[Trager et al. (2000)]{Trager:00} 
Trager, S.~C., Faber, S.~M., Worthey, G. \&
Gonz\'alez, J.~J. 2000, AJ, 119, 1645

\bibitem[Treu et al. (2010)]{Treu:10}
Treu T., Auger M.~W., Koopmans L.~V.~E., Gavazzi R., 
Marshall P.~J., Bolton A.~S., 2010, ApJ, 709, 1195  

\bibitem[Trevisan et al. (2012)]{Trevisan:12} 
Trevisan, M., Ferreras, I., de la Rosa, I.~G.,
La Barbera, F. \& de Carvalho, R.~R. 
2012, ApJl, 752, L27

\bibitem[Valdes et al. (2004)]{IndoUs}
Valdes, F., Gupta, R., Rose, J.,~A., Singh, H.~P., 
Bell, D.~J., 2004, ApJS, 152, 251

\bibitem[Van~Dokkum \& Conroy (2010)]{VdKConroy:10}
Van Dokkum, P.~G. \& Conroy, C. 2010, Nature, 468, 940

\bibitem[Van~Dokkum \& Conroy (2012)]{VdKConroy:12}
Van Dokkum, P.~G. \& Conroy, C. 2012, arXiv:1205.6473

\bibitem[Vazdekis et al. (1996)]{vaz:96}
Vazdekis, A., Casuso, E., Peletier, R.~F. \& Beckman, J.~E.,
1996, ApJs, 106, 307 

\bibitem[Vazdekis et al. (2003)]{vaz:03}
Vazdekis, A., Cenarro, A.~J., Gorgas, J., Cardiel, N. \&
Peletier, R.~F., 2003, MNRAS, 340, 1317

\bibitem[Vazdekis et al. (2010)]{MILES:10}
Vazdekis, A., S\'anchez-Bl'azquez, P., Falc\'on-Barroso, J., Cenarro, A.~J., 
Beasley, M.~A., Cardiel, N., Gorgas, J., Peletier, R.~F., 2010, MNRAS, 404, 1639

\bibitem[Vazdekis et al. (2012)]{MIUSCAT:12}
Vazdekis, A., Ricciardelli, E., Cenarro, A.~J., Rivero-Gonz\'alez, J.~G., 
D\'\i az-Garc\'\i a, L.~A. \& Falc\'on-Barroso, J.
2012, MNRAS, 424, 157

\bibitem[Worthey (1998)]{Worthey:98}
Worthey, G., 1998, PASP, 110, 888

\end{thebibliography}
\end{document}